\documentclass{esannV2}
\usepackage[dvips]{graphicx}
\usepackage[latin1]{inputenc}
\usepackage{amssymb,amsmath,array}
\usepackage[T1]{fontenc} 

\usepackage[pdfusetitle]{hyperref}
\hypersetup{colorlinks=true}

\usepackage{graphicx}
\usepackage{amsmath}
\usepackage{amssymb}
\usepackage{xargs}
\usepackage{xcolor}
\usepackage{subfig}
\usepackage{cleveref}
\usepackage{xspace}
\usepackage{url}

\newcommand{\ie}{\emph{i.e.}\xspace}
\newcommand{\eg}{\emph{e.g.}\xspace}
\newcommand{\wrt}{\emph{w.r.t.}\xspace}

\newcommand{\sorbetto}{\textsc{Sorbetto}\xspace}

\newcommand{\mysection}[1]{\vspace{2pt}\noindent\textbf{#1}}

\graphicspath{{images/}}

\begin{document}
\newcommand{\paperA}{paper~A~\cite{Pierard2024Foundations-arxiv}\xspace}
\newcommand{\paperB}{paper~B~\cite{Pierard2024TheTile-arxiv}\xspace}
\newcommand{\paperC}{paper~C~\cite{Halin2024AHitchhikers-arxiv}\xspace}
\newcommand{\PaperA}{Paper~A~\cite{Pierard2024Foundations-arxiv}\xspace}
\newcommand{\PaperB}{Paper~B~\cite{Pierard2024TheTile-arxiv}\xspace}
\newcommand{\PaperC}{Paper~C~\cite{Halin2024AHitchhikers-arxiv}\xspace}

\global\long\def\sampleSpace{\Omega}%
\global\long\def\aSample{\omega}%
\global\long\def\eventSpace{\Sigma}%
\global\long\def\anEvent{E}%
\global\long\def\measurableSpace{(\sampleSpace,\eventSpace)}%
\global\long\def\expectedValueSymbol{\mathbf{E}}%

\global\long\def\aPerformance{P}%
\global\long\def\allPerformances{\mathbb{\aPerformance}_{\measurableSpace}}%
\global\long\def\aSetOfPerformances{\Pi}%
\global\long\def\randVarSatisfaction{S}%
\global\long\def\aScore{X}%
\global\long\def\allScores{\mathbb{\aScore}_{\measurableSpace}}%
\newcommandx\domainOfScore[1][usedefault, addprefix=\global, 1=\aScore]{\mathrm{dom}(#1)}%
\global\long\def\evaluation{\mathrm{eval}}%
\global\long\def\opFilter{\mathrm{filter}_\randVarImportance}%
\global\long\def\opNoSkill{\mathrm{no\text{\textendash{}}skill}}%
\global\long\def\opPriorShift{\mathrm{shift}_{\pi\rightarrow\pi'}}%
\global\long\def\opChangePredictedClass{\mathrm{change}_{\randVarPredictedClass}}%
\global\long\def\opChangeGroundtruthClass{\mathrm{change}_{\randVarGroundtruthClass}}%
\global\long\def\opSwapGroundtruthAndPredictedClasses{\mathrm{swap}_{\randVarGroundtruthClass\leftrightarrow\randVarPredictedClass}}%
\global\long\def\opSwapClasses{\mathrm{swap}_{\classNeg\leftrightarrow\classPos}}%
\global\long\def\allWorstPerformances{\frownie}
\global\long\def\allBestPerformances{\smiley}

\global\long\def\randVarGroundtruthClass{Y}%
\global\long\def\randVarPredictedClass{\hat{Y}}%
\global\long\def\allClasses{\mathbb{C}}%
\global\long\def\aClass{c}%
\global\long\def\classNeg{c_-}%
\global\long\def\classPos{c_+}%
\global\long\def\sampleTN{tn}%
\global\long\def\sampleFP{fp}%
\global\long\def\sampleFN{fn}%
\global\long\def\sampleTP{tp}%
\global\long\def\eventTN{\{\sampleTN\}}%
\global\long\def\eventFP{\{\sampleFP\}}%
\global\long\def\eventFN{\{\sampleFN\}}%
\global\long\def\eventTP{\{\sampleTP\}}%
\global\long\def\scorePTN{PTN}%
\global\long\def\scorePFP{PFP}%
\global\long\def\scorePFN{PFN}%
\global\long\def\scorePTP{PTP}%
\global\long\def\scoreAccuracy{A}%
\global\long\def\scoreExpectedSatisfaction{\aScore_{\randVarSatisfaction}}%
\global\long\def\scoreTNR{TNR}%
\global\long\def\scoreFPR{FPR}%
\global\long\def\scoreTPR{TPR}%
\global\long\def\scoreFNR{FNR}%
\global\long\def\scoreNPV{NPV}%
\global\long\def\scoreFOR{FOR}%
\global\long\def\scorePPV{PPV}%
\global\long\def\scorePrecision{\scorePPV}%
\global\long\def\scoreFDR{FDR}%
\global\long\def\scoreJaccardNeg{J_-}%
\global\long\def\scoreJaccardPos{J_+}%
\global\long\def\scoreCohenKappa{\kappa}%
\global\long\def\scoreScottPi{\pi}%
\global\long\def\scoreFleissKappa{\kappa}%
\global\long\def\scoreBalancedAccuracy{BA}%
\global\long\def\scoreWeightedAccuracy{WA}%
\global\long\def\scoreYoudenJ{J_Y}
\global\long\def\scorePLR{PLR}%
\global\long\def\scoreNLR{NLR}%
\global\long\def\scoreOR{OR}%
\global\long\def\scoreSNPV{SNPV}%
\global\long\def\scoreSPPV{SPPV}%
\global\long\def\scoreACP{ACP}%
\global\long\def\scoreFOne{F_{1}}%
\newcommandx\scoreFBeta[1][usedefault, addprefix=\global, 1=\beta]{F_{#1}}%
\global\long\def\scoreFOne{\scoreFBeta[1]}%
\global\long\def\priorpos{\pi_+}%
\global\long\def\priorneg{\pi_-}%
\global\long\def\scoreBiasIndex{BI}%
\global\long\def\ratepos{\tau_+}%
\global\long\def\rateneg{\tau_-}%
\global\long\def\scoreACP{ACP}%
\global\long\def\scorePFour{P_4}%
\global\long\def\normalizedConfusionMatrix{C}%
\global\long\def\scoreConfusionMatrixDeterminant{|\normalizedConfusionMatrix|}

\global\long\def\allEntities{\mathbb{E}}%
\global\long\def\entitiesToRank{\mathbb{E}}%
\global\long\def\anEntity{\epsilon}%
\global\long\def\randVarImportance{I}%
\global\long\def\randVarCanonicalImportance{\randVarImportance_{a,b}}
\global\long\def\canonicalRankingScore{\rankingScore[\randVarCanonicalImportance]}
\newcommandx\rankingScore[1][usedefault, addprefix=\global, 1=\randVarImportance]{R_{#1}}%
\global\long\def\canonicalRankingScore{\rankingScore[\randVarImportance_{a,b}]}
\global\long\def\scoreVUT{VUT}%
\global\long\def\tileCurvePriors{\gamma_\pi}
\global\long\def\tileCurveRates{\gamma_\tau}
\global\long\def\relWorseOrEquivalent{\lesssim}%
\global\long\def\relBetterOrEquivalent{\gtrsim}%
\global\long\def\relEquivalent{\sim}%
\global\long\def\relBetter{>}%
\global\long\def\relWorse{<}%
\global\long\def\relIncomparable{\not\lesseqqgtr}%
\global\long\def\rank{\mathrm{rank}_\entitiesToRank}%
\global\long\def\ordering{\relWorseOrEquivalent}%
\global\long\def\invertedOrdering{\relBetterOrEquivalent}%

\global\long\def\LScityscapes{\ding{171}}
\global\long\def\LSade{\ding{170}}
\global\long\def\LSvoc{\ding{169}}
\global\long\def\LScoco{\ding{168}}

\global\long\def\indicatorSymbol{\mathbf{1}}
\global\long\def\realNumbers{\mathbb{R}}%
\global\long\def\aRelation{\mathcal{R}}%
\global\long\def\achievableByCombinations{\Phi}%
\global\long\def\allConvexCombinations{\mathrm{conv}}%
\newcommand{\indep}{\perp \!\!\! \perp}


\global\long\def\cityscapes{\LScityscapes{}~Cityscapes}
\global\long\def\ade{\LSade{}~ADE20K}
\global\long\def\voc{\LSvoc{}~Pascal VOC 2012}
\global\long\def\coco{\LScoco{}~COCO-Stuff 164k}

\newcommand{\MethodDesigner}{Bernadette}
\newcommand{\Benchmarker}{Leonard}
\newcommand{\AppDeveloper}{Howard}
\newcommand{\TheoreticalAnalyst}{Sheldon}

\newcommand{\tile}{Tile\xspace}
\newcommand{\tiles}{Tiles\xspace}
\newcommand{\valueTile}{Value Tile\xspace}
\newcommand{\valueTiles}{Value Tiles\xspace}
\newcommand{\baselineTile}{Baseline Value Tile\xspace}
\newcommand{\baselineTiles}{Baseline Value Tiles\xspace}
\newcommand{\SOTATile}{State-of-the-Art Value Tile\xspace}
\newcommand{\SOTATiles}{State-of-the-Art Value Tiles\xspace}
\newcommand{\noSkillTile}{No-Skill Tile\xspace}
\newcommand{\noSkillTiles}{No-Skill Tiles\xspace}
\newcommand{\skillTile}{Relative-Skill Tile\xspace}
\newcommand{\skillTiles}{Relative-Skill Tiles\xspace}
\newcommand{\correlationTile}{Correlation Tile\xspace}
\newcommand{\correlationTiles}{Correlation Tiles\xspace}
\newcommand{\rankingTile}{Ranking Tile\xspace}
\newcommand{\rankingTiles}{Ranking Tiles\xspace}
\newcommand{\entityTile}{Entity Tile\xspace}
\newcommand{\entityTiles}{Entity Tiles\xspace}

\global\long\def\aNonSkilledPerformance{\aPerformance_{\indep}}
\global\long\def\allNonSkilledPerformances{\mathbb{\aPerformance}^{\randVarGroundtruthClass\indep\randVarPredictedClass}_{\measurableSpace}}%

\global\long\def\allPriorFixedPerformances{\mathbb{\aPerformance}^{\priorpos}_{\measurableSpace}}%

\newcommand{\comma}{\,,}
\newcommand{\point}{\,.}

\newcommandx\unconditionalProbabilisticScore[1]{\aScore_{#1}^{U}}%
\global\long\def\formulaPTN{\unconditionalProbabilisticScore{\eventTN}}%
\global\long\def\formulaPFP{\unconditionalProbabilisticScore{\eventFP}}%
\global\long\def\formulaPFN{\unconditionalProbabilisticScore{\eventFN}}%
\global\long\def\formulaPTP{\unconditionalProbabilisticScore{\eventTP}}%
\global\long\def\formulapriorneg{\unconditionalProbabilisticScore{\{\sampleTN,\sampleFP\}}}%
\global\long\def\formulapriorpos{\unconditionalProbabilisticScore{\{\sampleFN,\sampleTP\}}}%
\global\long\def\formularateneg{\unconditionalProbabilisticScore{\{\sampleTN,\sampleFN\}}}%
\global\long\def\formularatepos{\unconditionalProbabilisticScore{\{\sampleFP,\sampleTP\}}}%
\global\long\def\formulaAccuracy{\unconditionalProbabilisticScore{\{\sampleTN,\sampleTP\}}}%

\newcommandx\conditionalProbabilisticScore[2]{\aScore_{#1 \vert #2}^{C}}%
\global\long\def\formulaTNR{\conditionalProbabilisticScore{\{\sampleTN\}}{\{\sampleTN,\sampleFP\}}}%
\global\long\def\formulaTPR{\conditionalProbabilisticScore{\{\sampleTP\}}{\{\sampleFN,\sampleTP\}}}%
\global\long\def\formulaNPV{\conditionalProbabilisticScore{\{\sampleTN\}}{\{\sampleTN,\sampleFN\}}}%
\global\long\def\formulaPPV{\conditionalProbabilisticScore{\{\sampleTP\}}{\{\sampleFP,\sampleTP\}}}%
\global\long\def\formulaJaccardNeg{\conditionalProbabilisticScore{\{\sampleTN\}}{\{\sampleTN,\sampleFP,\sampleFN\}}}%
\global\long\def\formulaJaccardPos{\conditionalProbabilisticScore{\{\sampleTP\}}{\{\sampleFP,\sampleFN,\sampleTP\}}}%

\global\long\def\scoreBennettS{S}


\global\long\def\allDomains{\mathbb{D}}%
\global\long\def\aDomain{d}%
\newcommandx\weightOfDomain[1][usedefault, addprefix=\global, 1=\aDomain]{\lambda_{#1}}%

\newcommand{\fourDomains}{\emph{easiest}, \emph{most difficult}, \emph{preponderant}, and \emph{bottleneck} domains\xspace}

\title{Multi-domain performance analysis\\with scores tailored to user preferences}

\author{S\'ebastien Pi\'erard, Adrien Deli\`ege, and Marc Van Droogenbroeck
%
\thanks{
    S\'ebastien Pi\'erard is funded by the Walloon region (Service Public de Wallonie Recherche, Belgium) under grant 2010235 - \emph{ARIAC by DIGITALWALLONIA4.AI}.
   Adrien Deli{\`e}ge is a \href{https://www.frs-fnrs.be}{F.R.S.-FNRS} postdoctoral researcher.
}
%
\vspace{.3cm}\\
Montefiore Institute, University of Li\`ege, Li\`ege, Belgium
}


\maketitle

\begin{abstract}
    The performance of algorithms, methods, and models tends to depend heavily on the distribution of cases on which they are applied, this distribution being specific to the applicative domain. 
    After performing an evaluation in several domains, it is highly informative to compute a (weighted) mean performance and, as shown in this paper, to scrutinize what happens during this averaging. To achieve this goal, we adopt a probabilistic framework and consider a performance as a probability measure (\eg, a normalized confusion matrix for a classification task). It appears that the corresponding weighted mean is known to be the summarization, and that only some remarkable scores assign to the summarized performance a value equal to a weighted arithmetic mean of the values assigned to the domain-specific performances. These scores include the family of ranking scores, a continuum parameterized by user preferences, and that the weights to consider in the arithmetic mean depend on the user preferences. Based on this, we rigorously define four domains, named easiest, most difficult, preponderant, and bottleneck domains, as functions of user preferences. After establishing the theory in a general setting, regardless of the task, we develop new visual tools for two-class classification.  
\end{abstract}

\section{Introduction}

The development of algorithms, methods, and models is often complicated by two types of uncertainty. First, the \emph{applicative domain}, and thus the distribution of inputs, is rarely unique and precisely known during development. Second, \emph{user preferences} regarding tradeoffs affecting the achievable performance (\eg, to what degree do they prefer false negatives to false positives in classification?) are also generally not precisely known and can differ from one user to another.

Developers can, of course, implement various techniques (continual learning, test-time adaptation, domain adaptation \cite{Wilson2020ASurvey}, domain generalization \cite{Gulrajani2020InSearch-arxiv}, \ldots) to make their \emph{entity} ---the term used in this paper for algorithm, method, or model--- usable in various domains, but the achievable performance still varies from one domain to another. After evaluating the developed entity in several domains, the developer is left with many questions, such as: Which domains are the strengths and weaknesses of the developed entities? Which domains have the greatest influence on average performance? Which domains constitute the bottlenecks for the average performance improvement? We need to address these questions from the end-user preferences perspective.

The objective of this paper is to rigorously demonstrate that, no matter what the task is, by taking an adequate way of averaging the domain-specific performances (\ie, the \emph{summarization} introduced in \cite{Pierard2020Summarizing}) as well as an adequate family of scores (\ie, the \emph{ranking scores} introduced in \cite{Pierard2025Foundations}), the \emph{easiest domain}, \emph{most difficult domain}, \emph{preponderant domain}, and \emph{bottleneck domain} become all well-defined and relative to some easy to interpret user preferences.

Our contributions are threefold:
\newline 
\textbf{1. Theoretical contribution on scores.} We highlight the fact that the \emph{expected value scores} (including all \emph{unconditional probabilistic scores}) and the \emph{expected value ratio scores} (including all \emph{probabilistic conditional scores} and all \emph{ranking scores}) have the remarkable property that the value taken for the averaged performance obtained by the \emph{summarization} technique~\cite{Pierard2020Summarizing} is equal to some weighted arithmetic mean of the values taken for the domain-specific performances, and provide closed-form formulas to compute these weights.
\newline  
\textbf{2. Contribution on user-centered multi-domain performance analysis.} Capitalizing on the remarkable property and focusing on the family of \emph{ranking scores} \cite{Pierard2025Foundations}, we rigorously define the \fourDomains as functions of a random variable that can be used to encode some user preferences for any task.
\newline     
\textbf{3. Contribution on visualization tools.} In the case of two-class crisp classification, we introduce new variants of the \tile, called \emph{flavors} \cite{Pierard2024TheTile-arxiv,Halin2024AHitchhikers-arxiv}, that allow us to visualize in a square how the \fourDomains depend on the user preferences.

\mysection{Notations.} Following \cite{Pierard2020Summarizing,Pierard2025Foundations,Halin2024AHitchhikers-arxiv}, we consider a probabilistic framework. A performance is a probability measure (\eg, a normalized confusion matrix for a classification task) on a measurable space $\measurableSpace$, where $\sampleSpace$ is the sample space and $\eventSpace$ the event space. The set of all performances is denoted by $\allPerformances$. We denote the set of domains by $\allDomains$, the performance in the domain $\aDomain$ by $\aPerformance_\aDomain$, and the weight arbitrarily given to the domain $\aDomain$ by $\weightOfDomain$. 
\section{Averaging performances with the summarization}

\newcommandx\summarizationWeight[1][usedefault, addprefix=\global, 1=\aScore]{\omega_{#1,\aDomain}}%

In this paper, we build upon the summarization \cite{Pierard2020Summarizing}, which is a probabilistic performance averaging technique that preserves the relationships between scores and preserves the probabilistic meaning of both unconditional and conditional probabilistic scores. It is applicable regardless of the task. The summarized performance $\overline{\aPerformance}$ is the following mixture of performances: 

\begin{equation}
    \overline{\aPerformance} = \frac{
        \sum_{\aDomain\in\allDomains} \weightOfDomain \aPerformance_{\aDomain}
    }{
        \sum_{\aDomain\in\allDomains} \weightOfDomain
    } 
    \qquad
    \textrm{with}
    \qquad
    \weightOfDomain\ge 0 \, \forall \aDomain \in \allDomains
    \point
    \label{eq:summarization}
\end{equation}
As noticed in~\cite{Pierard2020Summarizing}, for some remarkable scores $\aScore$, the summarization leads to
\begin{equation}
    \aScore(\overline{\aPerformance}) = \sum_{\aDomain\in\allDomains} \summarizationWeight \aScore(\aPerformance_{\aDomain})
    \qquad
    \textrm{with}
    \qquad
    \begin{cases}
        \summarizationWeight \ge 0 \, \forall \aDomain \in \allDomains\\
        \sum_{\aDomain\in\allDomains} \summarizationWeight = 1
    \end{cases}
    \point
    \label{eq:summarization-averaging-of-scores}
\end{equation}
We call $\weightOfDomain$ the \emph{domain weight} and $\summarizationWeight$ the \emph{summarization weight}.

\paragraph{Formulas for the weights for expected value scores.}

\newcommandx\expectedValueScore[1]{\aScore_{#1}^{EV}}%

We parameterize these scores by a random variable $V$ and define them as $\expectedValueScore{V}:\allPerformances\rightarrow 
\realNumbers:\aPerformance\mapsto\expectedValueScore{V}(\aPerformance)=\expectedValueSymbol_\aPerformance[V]$, where $\expectedValueSymbol$ is the mathematical expectation. With these scores, \cref{eq:summarization-averaging-of-scores} holds and the summarization weights do not depend on the performance:
\begin{equation}
    \summarizationWeight[\expectedValueScore{V}] = \frac{
        \weightOfDomain
    }{
        \sum_{\aDomain\in\allDomains} \weightOfDomain
    } \point
    \label{eq:summarization-weights-for-expected-value-scores}
\end{equation}

\paragraph{Formulas for the weights for expected value ratio scores.}

\newcommandx\expectedValueRatioScore[2]{\aScore_{#1,#2}^{EVR}}%

We parameterize these scores by two random variables, $V_1$ and $V_2\ne0$, and define them as $\expectedValueRatioScore{V_1}{V_2}:\domainOfScore[\expectedValueRatioScore{V_1}{V_2}]\rightarrow\realNumbers:\aPerformance\mapsto \expectedValueRatioScore{V_1}{V_2}(\aPerformance)=\frac{
    \expectedValueSymbol_{\aPerformance}[V_1]
}{
    \expectedValueSymbol_{\aPerformance}[V_2]
}$, where $\domainOfScore[\expectedValueRatioScore{V_1}{V_2}]=\{\aPerformance\in\allPerformances:\expectedValueSymbol_{\aPerformance}[V_2]\ne0\}$. With these scores, \cref{eq:summarization-averaging-of-scores} also holds, but contrary to the expected value scores, the summarization weights depend on the performance:
\begin{equation}    \summarizationWeight[\expectedValueRatioScore{V_1}{V_2}] = \frac{
        \weightOfDomain \expectedValueScore{V_2}(\aPerformance_\aDomain)
    }{
        \sum_{\aDomain\in\allDomains} \weightOfDomain \expectedValueScore{V_2}(\aPerformance_\aDomain)
    }
    = \frac{
        \weightOfDomain
    }{
        \sum_{\aDomain\in\allDomains} \weightOfDomain
    }
    \frac{
        \expectedValueScore{V_2}(\aPerformance_\aDomain)
    }{
        \expectedValueScore{V_2}(\overline{\aPerformance})
    }
    \point
    \label{eq:summarization-weights-for-expected-value-ratio-scores}
\end{equation}
\Cref{eq:summarization-weights-for-expected-value-scores,eq:summarization-weights-for-expected-value-ratio-scores} generalize the ones given in~\cite{Pierard2020Summarizing}, in the particular case of \emph{unconditional} and \emph{conditional} probabilistic scores, respectively.

\section{Taking user preferences into account through scores}

\long\def\easiestDomain{\mathrm{easiest\textrm{-}domain}(\randVarImportance)}
\long\def\mostDifficultDomain{\mathrm{most\textrm{-}difficult\textrm{-}domain}(\randVarImportance)}
\long\def\preponderantDomain{\mathrm{preponderant\textrm{-}domain}(\randVarImportance)}
\long\def\bottleneckDomain{\mathrm{bottleneck\textrm{-}domain}(\randVarImportance)}

In addition, we also build upon the ranking scores \cite{Pierard2025Foundations}, which establish a connection between the performances $\aPerformance$, the task (modeled through a random variable $\randVarSatisfaction$ called \emph{satisfaction}), and some user preferences (modeled through a non-negative random variable $\randVarImportance$ called \emph{importance}). These scores induce meaningful performance orderings~\cite{Pierard2025Foundations}, in the sense that these orderings satisfy the fundamental axioms of performance-based ranking. They are defined as: 
\begin{equation}
    \rankingScore:\domainOfScore[\rankingScore]\rightarrow\realNumbers:
    \aPerformance\mapsto
    \rankingScore(\aPerformance)=\frac{
        \expectedValueSymbol_{\aPerformance}[\randVarImportance\randVarSatisfaction]
    }{
        \expectedValueSymbol_{\aPerformance}[\randVarImportance]
    } \comma
    \label{eq:ranking-scores}
\end{equation}
where $\domainOfScore[\rankingScore]=\{\aPerformance\in\allPerformances:\expectedValueSymbol_{\aPerformance}[\randVarImportance]\ne0\}$. As $\rankingScore=\expectedValueRatioScore{\randVarImportance\randVarSatisfaction}{\randVarImportance}$, all ranking scores are particular cases of expected value ratio scores and \cref{eq:summarization-weights-for-expected-value-ratio-scores} also holds for them:
\begin{equation}
    \summarizationWeight[\rankingScore] = \frac{
        \weightOfDomain \expectedValueScore{\randVarImportance}(\aPerformance_\aDomain)
    }{
        \sum_{\aDomain\in\allDomains} \weightOfDomain \expectedValueScore{\randVarImportance}(\aPerformance_\aDomain)
    } = \frac{
        \weightOfDomain \expectedValueSymbol_{\aPerformance_\aDomain}[\randVarImportance]
    }{
        \sum_{\aDomain\in\allDomains} \weightOfDomain \expectedValueSymbol_{\aPerformance_\aDomain}[\randVarImportance]
    } = \frac{
        \weightOfDomain
    }{
        \sum_{\aDomain\in\allDomains} \weightOfDomain
    }
    \frac{
        \expectedValueSymbol_{\aPerformance_\aDomain}[\randVarImportance]
    }{
        \expectedValueSymbol_{\overline{\aPerformance}}[\randVarImportance]
    } \point
    \label{eq:summarization-weights-for-ranking-scores}
\end{equation}

Thanks to these scores, we can now analyze, in the perspective of user preferences, the performance of an entity that has been evaluated in several domains.

For a given entity, we can rank the domains, from the easiest (with the best performance) to the most difficult (with the worst performance). This ranking depends on the user preferences as follows:
\begin{equation}
    \easiestDomain
    =
    \underset{\aDomain \in \allDomains}{\arg \max} \, \rankingScore(\aPerformance_{\aDomain}) \comma
\end{equation}
\begin{equation}
    \mostDifficultDomain
    =
    \underset{\aDomain \in \allDomains}{\arg \min} \, \rankingScore(\aPerformance_{\aDomain}) \point
\end{equation}

We can also determine the preponderant domain \wrt user preferences. It is the domain for which the summarization weight is the largest:
\begin{equation}
    \preponderantDomain
    =
    \underset{\aDomain \in \allDomains}{\arg \max} \, \frac{
        \weightOfDomain \expectedValueSymbol_{\aPerformance_\aDomain}[\randVarImportance]
    }{
        \sum_{\aDomain\in\allDomains} \weightOfDomain \expectedValueSymbol_{\aPerformance_\aDomain}[\randVarImportance]
    }
    = \underset{\aDomain \in \allDomains}{\arg \max} \weightOfDomain \expectedValueSymbol_{\aPerformance_\aDomain}[\randVarImportance] \point
\end{equation}

A common practice in software engineering is to iteratively identify the parts that are bottlenecks and improve these parts. Similarly, if the objective is to design an entity that performs the best on average, one must be able to identify the domains that are bottlenecks. These are the domains for which the summarization weights are high and the values taken by the ranking scores are weak. We unify these two---sometimes contradictory---objectives by domain ablation: we define the bottleneck domain as the one that has to be removed from the set of domains to maximize a given ranking score:
\begin{equation}
   \bottleneckDomain
    =
    \underset{\aDomain \in \allDomains}{\arg \max} \, \rankingScore\left(
        \frac{
            \sum_{\aDomain'\in\allDomains\setminus \{\aDomain\}} \weightOfDomain[\aDomain'] \aPerformance_{\aDomain'}
        }{
            \sum_{\aDomain'\in\allDomains\setminus \{\aDomain\}} \weightOfDomain[\aDomain']
        } 
   \right)
   \point
\end{equation}

\section{Visualization tools for two-class crisp classification}

Lastly, we build upon the \tile \cite{Pierard2024TheTile-arxiv}, which is a tool for two-class crisp classification to visualize functions of the importance $\randVarImportance$ on a square space. The horizontal axis indicates the relative importance considered by the user for the \emph{true positives} \wrt the \emph{true negatives}. The vertical axis indicates the relative importance for the \emph{false negatives} \wrt the \emph{false positives}.

\newcommand{\normalizedConfusionMatrices}{
    \hspace{30mm}
    \begin{minipage}[b]{5cm}
        \begin{tabular}{|c|c|c|}
        \cline{2-3}
        \multicolumn{1}{c|}{$\aPerformance_{d_{1}}$} & $\randVarPredictedClass=\classNeg$ & $\randVarPredictedClass=\classPos$\tabularnewline
        \hline 
        $\randVarGroundtruthClass=\classNeg$ & $0.02$ & $0.12$\tabularnewline
        \hline 
        $\randVarGroundtruthClass=\classPos$ & $0.01$ & $0.85$\tabularnewline
        \hline 
        \end{tabular}
        
        \vspace{5mm}
        \begin{tabular}{|c|c|c|}
        \cline{2-3}
        \multicolumn{1}{c|}{$\aPerformance_{d_{2}}$} & $\randVarPredictedClass=\classNeg$ & $\randVarPredictedClass=\classPos$\tabularnewline
        \hline 
        $\randVarGroundtruthClass=\classNeg$ & $0.68$ & $0.08$\tabularnewline
        \hline 
        $\randVarGroundtruthClass=\classPos$ & $0.10$ & $0.14$\tabularnewline
        \hline 
        \end{tabular}
        
        \vspace{5mm}
        \begin{tabular}{|c|c|c|}
        \cline{2-3}
        \multicolumn{1}{c|}{$\aPerformance_{d_{3}}$} & $\randVarPredictedClass=\classNeg$ & $\randVarPredictedClass=\classPos$\tabularnewline
        \hline 
        $\randVarGroundtruthClass=\classNeg$ & $0.41$ & $0.19$\tabularnewline
        \hline 
        $\randVarGroundtruthClass=\classPos$ & $0.10$ & $0.30$\tabularnewline
        \hline 
        \end{tabular}
        
        \vspace{5mm}
        \begin{tabular}{|c|c|c|}
        \cline{2-3}
        \multicolumn{1}{c|}{$\overline{\aPerformance}$} & $\randVarPredictedClass=\classNeg$ & $\randVarPredictedClass=\classPos$\tabularnewline
        \hline 
        $\randVarGroundtruthClass=\classNeg$ & $0.37$ & $0.13$\tabularnewline
        \hline 
        $\randVarGroundtruthClass=\classPos$ & $0.07$ & $0.43$\tabularnewline
        \hline 
        \end{tabular}

        \vspace{50mm}
    \end{minipage}
}

\begin{figure}[t]
\resizebox{\linewidth}{!}{
    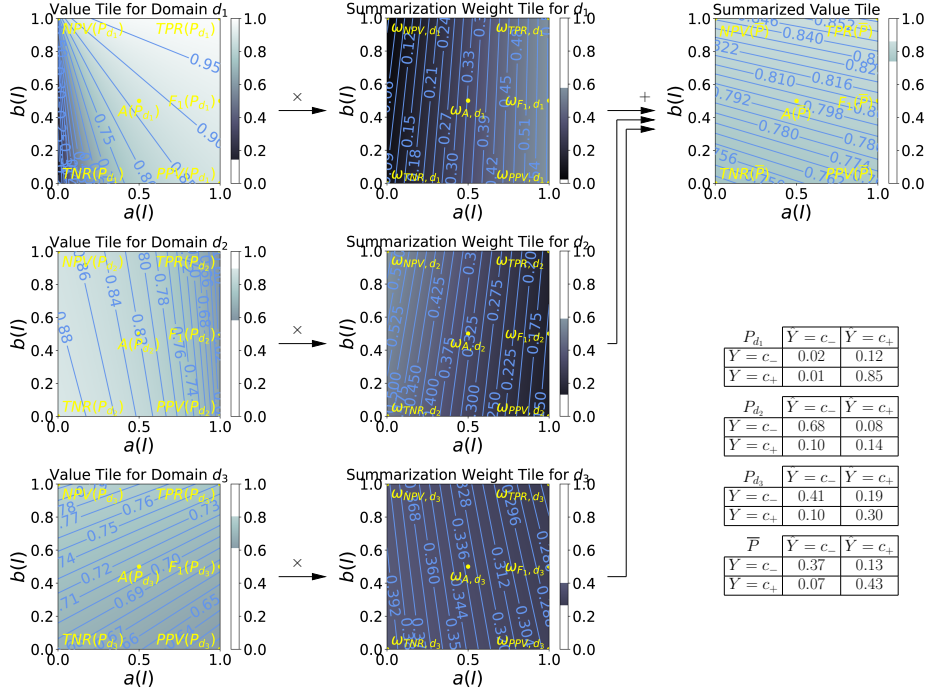
}

\caption{
    \tiles showing the summarization on 3 domains: by multiplying each domain-specific \emph{Value Tile} by the corresponding \emph{Summarization Weight Tile} and adding the results together, one obtains the \emph{Summarized Value Tile}, \ie the \emph{Value Tile} for $\overline{\aPerformance}$. This \tile is exactly the same as the \emph{Value Tile} that would be obtained from $\overline{\aPerformance}$ after computing it with \cref{eq:summarization}. However, by scrutinizing what happens during the performance averaging, it becomes clear that the actual weights to consider strongly depend on the user preferences (\ie, point in the \tile), which is something hidden in \cref{eq:summarization}.
    \label{fig:summarization}
}

\end{figure}

\paragraph{Defining the task.}
From the performance perspective~\cite{Pierard2025Foundations}, two-class crisp classification is defined by the sample space $\sampleSpace=\{\sampleTN,\sampleFP,\sampleFN,\sampleTP\}$ and the satisfaction $\randVarSatisfaction=\indicatorSymbol_{\{\sampleTN,\sampleTP\}}$.  
The samples $\sampleTN$, $\sampleFP$, $\sampleFN$, and $\sampleTP$ are traditionally interpreted as a true negative, false positive, false negative, and true positive, respectively.

\paragraph{Setting the user preferences.} In our setting, the user can freely specify some preferences through the importance values $\randVarImportance(\sampleTN)$, $\randVarImportance(\sampleFP)$, $\randVarImportance(\sampleFN)$, and $\randVarImportance(\sampleTP)$. The generic formula for the ranking scores given in \cref{eq:ranking-scores} is particularized as follows:
\begin{equation*}
    \rankingScore(\aPerformance)
    = \frac{
        \aPerformance(\{\sampleTN\}) \, \randVarImportance(\sampleTN)
        + \aPerformance(\{\sampleTP\}) \, \randVarImportance(\sampleTP)
    }{
        \aPerformance(\{\sampleTN\}) \, \randVarImportance(\sampleTN)
        + \aPerformance(\{\sampleFP\}) \, \randVarImportance(\sampleFP)
        + \aPerformance(\{\sampleFN\}) \, \randVarImportance(\sampleFN)
        + \aPerformance(\{\sampleTP\}) \, \randVarImportance(\sampleTP)
    }
    \point
\end{equation*}

\paragraph{Using the \tile.}
As the ranking scores sharing the same values for $a(\randVarImportance)=\frac{\randVarImportance(\sampleTP)}{\randVarImportance(\sampleTN)+\randVarImportance(\sampleTP)}$ and $b(\randVarImportance)=\frac{\randVarImportance(\sampleFN)}{\randVarImportance(\sampleFP)+\randVarImportance(\sampleFN)}$ induce the same ranking, 
it has been proposed in \cite{Pierard2024TheTile-arxiv} to focus on the \emph{canonical ranking scores} such that $\randVarImportance(\sampleTN) + \randVarImportance(\sampleTP) = \randVarImportance(\sampleFP) + \randVarImportance(\sampleFN)$ and to map them on the $(a,b)\in[0,1]^2$ space called \emph{\tile}. 
So, $\randVarImportance(\sampleTN)=1-a$, $\randVarImportance(\sampleFP)=1-b$, $\randVarImportance(\sampleFN)=b$, and $\randVarImportance(\sampleTP)=a$. 
Note that the accuracy, the true negative rate, the true positive rate, the negative predictive value, the positive predictive value, and $\scoreFOne$ are particular cases of canonical ranking scores.

\paragraph{Visualizing the summarization.}
\Cref{fig:summarization} depicts the summarization, \ie \cref{eq:summarization}, with \tiles. The values taken by the canonical ranking scores and the weights $\summarizationWeight[\rankingScore]$ are shown, respectively, on \emph{Value Tiles} and \emph{Summarization Weight Tiles}.

\paragraph{Introducing new \emph{flavors}.}
\begin{figure*}
\subfloat[easiest\label{fig:flavor-easiest}]{\includegraphics[width=0.23\linewidth]{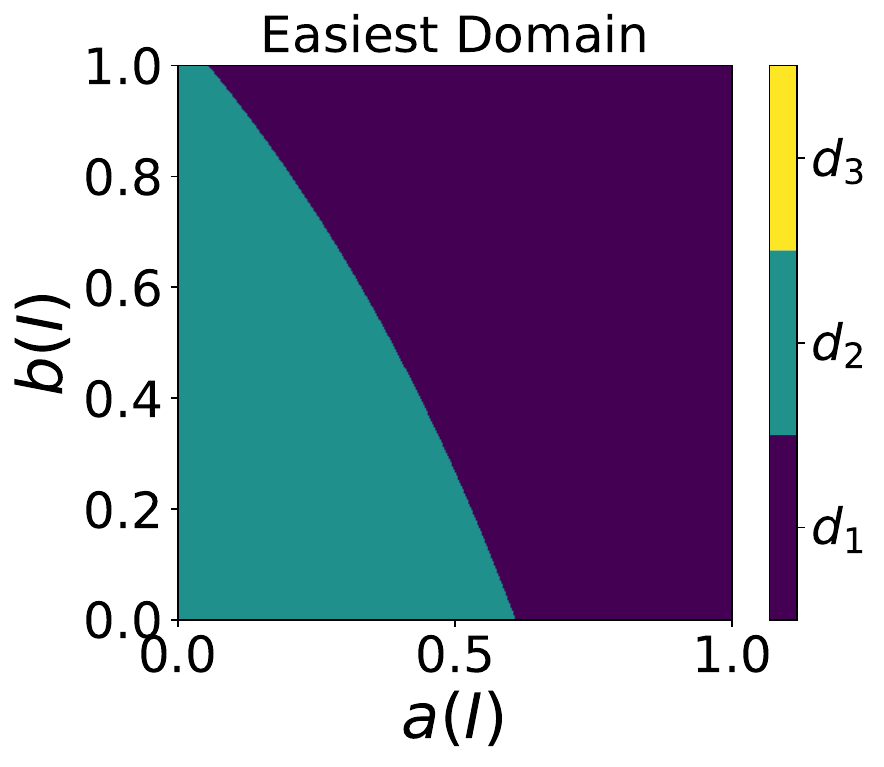}

}
\hfill{}
\subfloat[most difficult\label{fig:flavor-most-difficult}]{\includegraphics[width=0.23\linewidth]{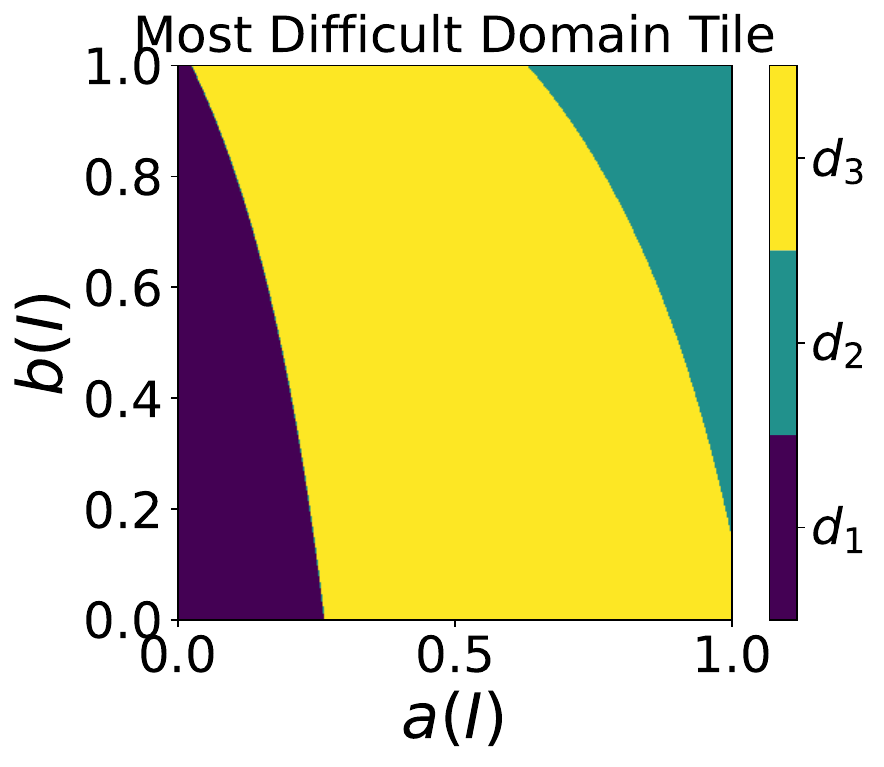}

}
\hfill{}
\subfloat[preponderant\label{fig:flavor-preponderant}]{\includegraphics[width=0.23\linewidth]{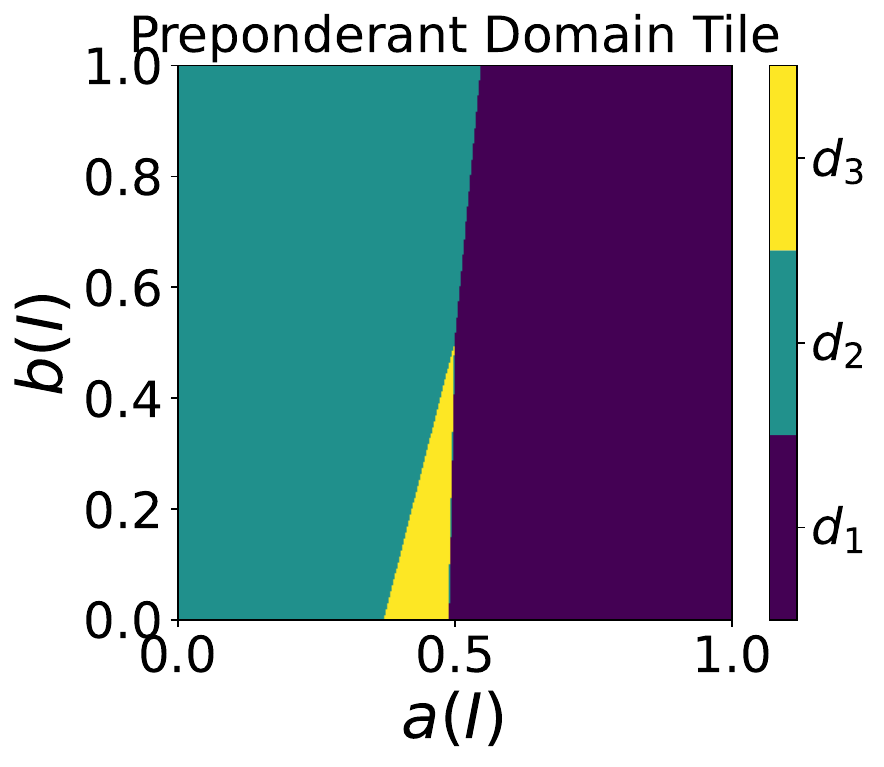}

}
\hfill{}
\subfloat[bottleneck\label{fig:flavor-bottleneck}]{\includegraphics[width=0.23\linewidth]{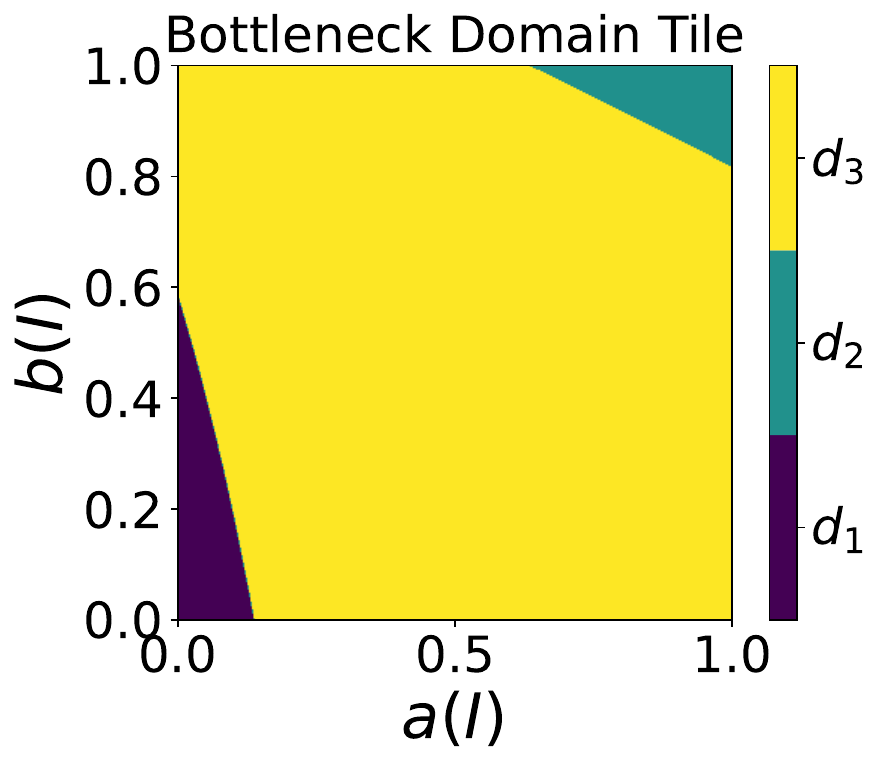}

}

\caption{
    In this work, we propose four new flavors for the \tile, as visual tools to perform multi-domain performance analyses \wrt user preferences. These are the \tiles obtained for the example of \cref{fig:summarization}. Notably, this analysis covers the two sources of uncertainty met in a model development: the domain dimension (the colors on the \tiles), and the user preferences (the position on the \tiles).
    \label{fig:new-tile-flavors}
}

\end{figure*}
We extend the catalog of flavors (\ie, functions of the importance $\randVarImportance$) compiled in \cite{Halin2024AHitchhikers-arxiv} and implemented in the \sorbetto library \cite{Pierard2025Sorbetto-esann}, and propose to depict on the \tile the easiest, the most difficult, the preponderant, and the bottleneck domains (see \cref{fig:new-tile-flavors}). Practically, to improve on the worst case scenario, the developer should improve on the most difficult domain, and to improve on average, improve on the bottleneck domain. Currently, only the central point is generally considered (\ie the accuracy), such as in~\cite{Gulrajani2020InSearch-arxiv}, overlooking entirely the dimension of user preferences.

\section{Conclusion}

By considering the technique for averaging domain-specific performances known as the \emph{summarization} \cite{Pierard2020Summarizing} and the family of scores known as the \emph{ranking scores} \cite{Pierard2025Foundations}, we rigorously defined the \fourDomains as functions of some user preferences. First, we did it in a way that makes it possible for our theoretical results to be used for any task. 
Then, we particularized our results to the task of two-class crisp classification, demonstrating how the dependence of the four defined domains regarding the user preferences can be visualized on a square space that is known as the \tile \cite{Pierard2024TheTile-arxiv}. By doing so, we extended the catalog of \tile flavors compiled in \cite{Halin2024AHitchhikers-arxiv} and currently implemented in the \sorbetto library \cite{Pierard2025Sorbetto-esann}.


\begin{footnotesize}

\end{footnotesize}


\end{document}